\documentstyle[12pt]{article}
\begin{document}

\small
\hoffset=-1truecm
\voffset=-2truecm
\title{\bf The Casimir effect at finite temperature in the presence of compactified universal extra dimensions}
\author{Hongbo Cheng\footnote {E-mail address:
hbcheng@public4.sta.net.cn}\\
Department of Physics, East China University of Science and
Technology,\\ Shanghai 200237, China}

\date{}
\maketitle

\begin{abstract}
In this paper the Casimir effect for parallel plates at finite
temperature in the presence of compactified universal extra
dimensions is analyzed. We show the thermal corrections to the
effect in detail. We investigate the Casimir effect for different
size of universal extra dimensions.
\end{abstract}
\vspace{8cm} \hspace{1cm}
PACS number(s): 11.10.Kk, 04.62.+v

\newpage

\noindent 1. \hspace{0.4cm}Introduction

More than 80 years ago the idea that our world has more than three
spatial dimensions was put forward by Kaluza and Klein [1, 2].
They introduced an additional compactified dimension in order to
unify gravity and classical electrodynamics. Recently, as one of
the fundamental aspects, string theory needs seven additional
spatial dimensions to unify the quantum mechanics and gravity. It
is interesting that the order of the compactification scale of the
extra dimensions is different in the branches of string theory.
Some models of string theory expect that the typical size of
compactified universal extra dimensions is of order $10^{-35}m$,
which means that observing the effect is beyond our experimental
reach of today and near future [3, 4]. In some approaches large
extra dimensions were invoked for solving the hierarchy problem
[5-8]. It is supposed that gauge fields are localized on a four
dimensional brane, our real universe, and only a graviton can
propagate in the extra space transverse to the brane.

The Casimir effect is a fundamental aspect of quantum field theory
in confined geometries and has been a subject of extensive
research [9-14]. The precision of the measurements has been
greatly improved experimentally [15]. Now the Casimir effect
becomes an experimentally powerful method for the study of new
physics beyond the standard model. There are many cosmological
uses of Casimir effect contributing with the right order of
magnitude to the observed value of the cosmological constant
[16-18]. A lot of topics related to the Casimir effect have been
explored in the context of string theory [19-21]. More progresses
of the effect were made to stabilize the extra dimensions [22,
23].

It is crucial but difficult to investigate how many dimensions our
universe has and its structure theoretically and experimentally.
If our spacetime has ten or more dimensions, we should observe the
new phenomena related to the existence of the additional
dimensions. Probing the possible existence and size of universal
extra dimensions (UXDs) by means of Casimir effect attracts more
attentions of the physical community [24, 25]. The Casimir force
between two parallel plates in the presence of UXDs is
investigated. The recent results show that the lower bound in
energy is 300 GeV, which corresponds to a maximum size of extra
dimension of about $10^{-9}nm$ [24]. There is an upper limit of
$R\leq10nm$ with one UXD [25].

Quantum field theory at finite temperature shares many of the
effects. Thermal influence on the general Casimir effect can not
be neglected in many cases [26-28]. It is essential to discuss the
Casimir effect for the system consisting of two parallel plates in
the presence of UXDs under a nonzero temperature environment. In
this letter we consider some important and interesting arguments
on the Casimir effect for the parallel plates at finite
temperature with UXDs in detail. We predict that the thermal
influence can not be omitted and the phenomena related to UXD is
more manifest.

At different temperature and with various order of compactified
extra dimensions of spacetime the description of the Casimir
effect for parallel plates change greatly. Here we discuss the
thermal corrections to the Casimir effect for the system by means
of the Kaluza-Klein theory and finite-temperature field theory [2,
3, 29, 30]. By regularizing the total energy, we find that the
sign of Casimir energy for parallel plates in the world with one
extra dimension is also positive at sufficiently high
temperatures. The plates will repulse each other for their large
enough distance. The phenomena related to Casimir effect with UXD
become manifest. We try to test the size of UXD suggested in
string theory. In this paper we derive the Casimir energy for
parallel plates at finite temperature with UXD. We discuss the
Casimir energy and Casimir force at different temperatures and
with various size of extra dimension. Finally the conclusions are
emphasized.

\vspace{0.8cm} \noindent 2.\hspace{0.4cm}The Casmir energy for
parallel plates at finite temperature in the spacetime with one
extra dimension

In the Kaluza-Klein (KK) approach the Lagrangian density of a
simple model of scalar field in (4+1)-dimensional spacetime takes
the form,

\begin{equation}
{\cal L}=\frac{1}{2}\partial_{A}\Phi\partial^{A}\Phi
\end{equation}

\noindent here $\Phi(x^{A})=\Phi(x^{\mu},y)$ is the field,
$A=0,1,2,3,5$ and $\mu=0,1,2,3$ belonging to four-dimensional
coordinates and $y$ to extra coordinate. According to the
compactification of extra dimension on a circle, the field $\Phi$
can be expanded in the harmonics as follow,

\begin{equation}
\Phi(x,y)=\sum_{n=-\infty}^{\infty}\phi_{n}(x)e^{iny/L}
\end{equation}

\noindent By substituting (2) into (1), the Lagrangian density (1)
becomes,

\begin{equation}
{\cal
L}=\frac{1}{2}\partial_{\mu}\phi_{0}\partial^{\mu}\phi_{0}+\sum_{k=1}^{\infty}(\partial_{\mu}\phi_{k}\partial^{\mu}\phi_{k}^{*}+\frac{k^{2}}{L^{2}}\phi_{k}\phi_{k}^{*})
\end{equation}

\noindent where $L$ is the radius of UXD.

In finite-temperature field theories the imaginary time formalism
can be used to describe the scalar fields in thermal equilibrium
[26-30]. We introduce a partition function for a system in the
presence of UXD,

\begin{equation}
Z=N\int_{period}\prod_{k}D\phi_{k}\exp[\int_{0}^{\beta}d\tau\int
d^{3}x\cal L(\phi_{k}, \partial_{E}\phi_{k})]
\end{equation}

\noindent where $\cal L$ is the Lagrangian density denoted as (3),
$N$ a constant and "period" means
$\phi_{k}(0,\mathbf{x})=\phi_{k}(\tau,\mathbf{x})$,
$k=0,1,2,\cdot\cdot\cdot$. $\beta=\frac{1}{T}$ is the inverse of
the temperature. The scalar fields $\phi_{k}$ satisfy the
Klein-Gordon equations,

\begin{equation}
(\partial_{\mu}\partial^{\mu}-\frac{k^{2}}{L^{2}})\phi_{k}(x)=0
\end{equation}

\noindent where $k=0,1,2,\cdot\cdot\cdot$. The field confining
between the two parallel plates satisfy the Dirichlet boundary
conditions $\phi_{k}(x)|_{\partial\Omega}=0$, $\partial\Omega$
positions of the plates. Following [26-28], the generalized zeta
function reads,

\begin{eqnarray}
\zeta(s;-\partial_{E})=Tr(-\partial_{E})^{-s}\hspace{5cm}\nonumber\\
=\int
d^{2}k\sum_{m=1}^{\infty}\sum_{l=-\infty}^{\infty}\sum_{n=0}^{\infty}[k^{2}+\frac{m^{2}\pi^{2}}{R^{2}}+\frac{n^{2}}{L^{2}}+(\frac{2l\pi}{\beta})^{2}]^{-s}
\end{eqnarray}

\noindent where $R$ is the distance of the plates and
$\partial_{E}=\frac{\partial^{2}}{\partial\tau^{2}}+\nabla^{2}$
with $\tau=it$. Furthermore, the function (6) can also be
expressed in terms of Epstein zeta function $E$ and Riemann zeta
function $\zeta$,

\begin{eqnarray}
\zeta(s;-\partial_{E})=\pi^{3-2s}\frac{\Gamma(s-1)\zeta(2s-2)}{\Gamma(s)}\frac{1}{R^{2-2s}}
+\pi\frac{\Gamma(s-1)}{\Gamma(s)}E_{2}(s-1;\frac{\pi^{2}}{R^{2}},\frac{1}{L^{2}})\nonumber\\
+2\pi\frac{\Gamma(s-1)}{\Gamma(s)}E_{2}(s-1;\frac{\pi^{2}}{R^{2}},\frac{4\pi^{2}}{\beta^{2}})
+2\pi\frac{\Gamma(s-1)}{\Gamma(s)}E_{3}(s-1;\frac{\pi^{2}}{R^{2}},\frac{1}{L^{2}},\frac{4\pi^{2}}{\beta^{2}})
\end{eqnarray}

\noindent the energy density of the model with thermal corrections
and UXDs is,

\begin{eqnarray}
\varepsilon=-\frac{\partial}{\partial\beta}(\frac{\partial\zeta(s;-\partial_{E})}{\partial
s}|_{s=0})\hspace{9cm}\nonumber\\
=-\frac{\pi^{\frac{7}{2}}}{2R^{3}}\Gamma(-\frac{3}{2})\zeta(-3)
-\frac{\sqrt{\pi}}{2}\Gamma(-\frac{3}{2})E_{2}(-\frac{3}{2};\frac{\pi^{2}}{R^{2}},\frac{1}{L^{2}})\hspace{5cm}\nonumber\\
+2^{\frac{3}{2}}\pi^{2}(\beta
R)^{-\frac{3}{2}}\sum_{n_{1},n_{2}=1}^{\infty}(\frac{n_{2}}{n_{1}})^{\frac{3}{2}}K_{\frac{3}{2}}(\beta
\frac{\pi}{R}n_{1}n_{2})\hspace{6cm}\nonumber\\
+(2\pi^{2})^{\frac{3}{2}}\beta^{-\frac{1}{2}}R^{-\frac{5}{2}}
\sum_{n_{1},n_{2}=1}^{\infty}n_{1}^{-\frac{1}{2}}n_{2}^{\frac{5}{2}}
[K_{\frac{1}{2}}(\beta\frac{\pi}{R}n_{1}n_{2})+K_{\frac{5}{2}}(\beta\frac{\pi}{R}n_{1}n_{2})]\hspace{2.5cm}\nonumber\\
+4\pi\sum_{k=0}^{\infty}\frac{8^{-k}(k+1)}{k!}\beta^{-k-2}\prod_{j=1}^{k}[9-(2j-1)^{2}]
\sum_{n_{1},n_{2},n_{3}=1}^{\infty}n_{1}^{-k-2}(\frac{\pi^{2}}{R^{2}}n_{2}^{2}+\frac{1}{L^{2}}n_{3}^{2})^{-\frac{k-1}{2}}\hspace{-0.5cm}\nonumber\\
\times\exp[-\beta
n_{1}(\frac{\pi^{2}}{R^{2}}n_{2}^{2}+\frac{1}{L^{2}}n_{3}^{2})^{\frac{1}{2}}] \hspace{4cm}\nonumber\\
+4\pi\sum_{k=0}^{\infty}\frac{8^{-k}}{k!}\beta^{-k-1}\prod_{j=1}^{k}[9-(2j-1)^{2}]
\sum_{n_{1},n_{2},n_{3}=1}^{\infty}n_{1}^{-k-1}(\frac{\pi^{2}}{R^{2}}n_{2}^{2}+\frac{1}{L^{2}}n_{3}^{2})^{-\frac{k-2}{2}}\hspace{0.5cm}\nonumber\\
\times\exp[-\beta
n_{1}(\frac{\pi^{2}}{R^{2}}n_{2}^{2}+\frac{1}{L^{2}}n_{3}^{2})^{\frac{1}{2}}]
\hspace{4cm}
\end{eqnarray}

\noindent where $K_{\nu}(z)$ is the modified Bessel function of
the second kind. Now we introduce two dimensionless variables, the
scaled temperature and ratio of plates distance to UXD size
respectively,

\begin{eqnarray}
\xi=TL \nonumber \\
\mu=\frac{R}{L}
\end{eqnarray}

\noindent By regularizing the expression, we obtain the Casimir
energy for parallel plates at nonzero temperature with UXD as
follow,

\begin{eqnarray}
\varepsilon_{C}=-\frac{1}{2}\Gamma(2)\zeta(4)\frac{1}{\mu^{3}}\frac{1}{L^{3}}
+\frac{\pi^{-3}}{4}\Gamma(2)\zeta(4)\frac{1}{L^{3}}
-\frac{\pi^{-\frac{9}{2}}}{4}\Gamma(\frac{5}{2})\zeta(5)\mu\frac{1}{L^{3}}\hspace{1cm}\nonumber\\
-\frac{1}{\mu}\frac{1}{L^{3}}\sum_{n_{1},n_{2}=1}^{\infty}(\frac{n_{2}}{n_{1}})^{2}K_{2}(2\mu n_{1}n_{2})
\hspace{5cm}\nonumber\\
+2^{\frac{3}{2}}\pi^{2}(\frac{\xi}{\mu})^{\frac{3}{2}}\frac{1}{L^{3}}
\sum_{n_{1},n_{2}=1}^{\infty}(\frac{n_{2}}{n_{1}})^{\frac{3}{2}}K_{\frac{3}{2}}(\frac{\pi
n_{1}n_{2}}{\xi\mu})\hspace{3.5cm}\nonumber\\
+(2\pi^{2})^{\frac{3}{2}}\xi^{\frac{1}{2}}\mu^{-\frac{5}{2}}\frac{1}{L^{3}}
\sum_{n_{1},n_{2}=1}^{\infty}n_{1}^{-\frac{1}{2}}n_{2}^{\frac{5}{2}}[K_{\frac{1}{2}}(\frac{\pi
n_{1}n_{2}}{\xi\mu})+K_{\frac{5}{2}}(\frac{\pi
n_{1}n_{2}}{\xi\mu})]\nonumber\\
+\frac{4\pi}{L^{3}}\sum_{k=0}^{\infty}\frac{8^{-k}(k+1)}{k!}\xi^{k+2}
\prod_{j=1}^{k}[9-(2j-1)^{2}]
\sum_{n_{1},n_{2},n_{3}=1}^{\infty}n_{1}^{-k-2}(\frac{\pi^{2}}{\mu^{2}}n_{2}^{2}+n_{3}^{2})^{-\frac{k-1}{2}}
\hspace{-2cm}\nonumber\\
\times\exp[-\frac{n_{1}}{\xi}(\frac{\pi^{2}}{\mu^{2}}n_{2}^{2}+n_{3}^{2})^{\frac{1}{2}}]\hspace{3cm}\nonumber\\
+\frac{4\pi}{L^{3}}\sum_{k=0}^{\infty}\frac{8^{-k}}{k!}\xi^{k+1}
\prod_{j=1}^{k}[9-(2j-1)^{2}]
\sum_{n_{1},n_{2},n_{3}=1}^{\infty}n_{1}^{-k-1}(\frac{\pi^{2}}{\mu^{2}}n_{2}^{2}+n_{3}^{2})^{-\frac{k-2}{2}}
\hspace{-1cm}\nonumber\\
\times\exp[-\frac{n_{1}}{\xi}(\frac{\pi^{2}}{\mu^{2}}n_{2}^{2}+n_{3}^{2})^{\frac{1}{2}}]\hspace{3cm}
\end{eqnarray}

\noindent The terms with series converge very quickly [14, 27] and
only the first several summands need to be taken into account for
numerical calculation to further discussions.

\vspace{0.8cm}
\noindent 3.\hspace{0.4cm} The Casimri effect for
parallel plates at finite temperature in the spacetime with on
extra dimension

At finite temperature in the cosmological background with one UXD,
we consider the sign of Casimir energy for parallel plates and the
nature of Casimir force between them to present the Casimir effect
clearly. The Casimir force is denoted as
$F_{C}=-\frac{1}{L}\frac{\partial\varepsilon_{C}}{\partial\mu}$.
We analyze the Casimir energy (10) in the limits. If the
temperature is high enough $\xi\gg1$, then

\begin{eqnarray}
\varepsilon_{C}(\xi\gg1)=2^{\frac{3}{2}}\pi^{2}(\frac{\xi}{\mu})^{\frac{3}{2}}\frac{1}{L^{3}}
\sum_{n_{1},n_{2}=1}^{\infty}(\frac{n_{2}}{n_{1}})^{\frac{3}{2}}K_{\frac{3}{2}}(\frac{\pi
n_{1}n_{2}}{\xi\mu})\hspace{4.5cm}\nonumber\\
+(2\pi^{2})^{\frac{3}{2}}\xi^{\frac{1}{2}}\mu^{-\frac{5}{2}}\frac{1}{L^{3}}
\sum_{n_{1},n_{2}=1}^{\infty}n_{1}^{-\frac{1}{2}}n_{2}^{\frac{5}{2}}[K_{\frac{1}{2}}(\frac{\pi
n_{1}n_{2}}{\xi\mu})+K_{\frac{5}{2}}(\frac{\pi
n_{1}n_{2}}{\xi\mu})]\hspace{1cm}\nonumber\\
+\frac{4\pi}{L^{3}}\sum_{k=0}^{\infty}\frac{8^{-k}(k+1)}{k!}\xi^{k+2}
\prod_{j=1}^{k}[9-(2j-1)^{2}]
\sum_{n_{1},n_{2},n_{3}=1}^{\infty}n_{1}^{-k-2}(\frac{\pi^{2}}{\mu^{2}}n_{2}^{2}+n_{3}^{2})^{-\frac{k-1}{2}}
\hspace{-1.5cm}\nonumber\\
\times\exp[-\frac{n_{1}}{\xi}(\frac{\pi^{2}}{\mu^{2}}n_{2}^{2}+n_{3}^{2})^{\frac{1}{2}}]\hspace{3cm}\nonumber\\
+\frac{4\pi}{L^{3}}\sum_{k=0}^{\infty}\frac{8^{-k}}{k!}\xi^{k+1}
\prod_{j=1}^{k}[9-(2j-1)^{2}]
\sum_{n_{1},n_{2},n_{3}=1}^{\infty}n_{1}^{-k-1}(\frac{\pi^{2}}{\mu^{2}}n_{2}^{2}+n_{3}^{2})^{-\frac{k-2}{2}}
\hspace{-1cm}\nonumber\\
\times\exp[-\frac{n_{1}}{\xi}(\frac{\pi^{2}}{\mu^{2}}n_{2}^{2}+n_{3}^{2})^{\frac{1}{2}}]\hspace{3cm}\nonumber\\
>0 \hspace{10cm}
\end{eqnarray}

\noindent For a definite temperature, when the plates distance is
much larger or less than the radius of UXD, the expression for
Casimir energy becomes,

\begin{equation}
\varepsilon_{C}(\mu\gg1)=-\frac{\Gamma(\frac{5}{2})\zeta(5)}{4\pi^{\frac{9}{2}}}\mu\frac{1}{L^{3}}
\end{equation}

\noindent and

\begin{equation}
\varepsilon_{C}(\mu\ll1)=-\frac{\Gamma(2)\zeta(4)}{2}\frac{1}{\mu^{3}}\frac{1}{L^{3}}
\end{equation}

\noindent respectively. We find that the Casimir energy is an
increasing function of the variable $\xi$. For a value of
temperature there exists a maximum of Casimir energy depending on
the ratio $\mu$. The Casimir energy is calculated from equation
(10). For some values of $\xi$, the numerical calculations lead to
the data presented in Figure 1. According to (10) the special
scaled temperature $\xi_{0}=0.0947$ is obtained.

If the temperature is chosen as $\xi<\xi_{0}$, the Casimir energy
keeps negative. At $\mu=\mu_{f}$, the energy is equal to the
maximum. Having solved equation (10), we obtain the relation
between the special ratio $\mu_{f}$ and temperature, with
$\mu_{f}$ growing as scaled temperature $\xi$ as shown in Figure
2. In the spacetime with one UXD, the two parallel plates will
attract each other if $\frac{R}{L}<\mu_{f}$, and if
$\frac{R}{L}>\mu_{f}$, the plates will remain repulsive.

If the temperature is sufficiently high like $\xi>\xi_{0}$, the
Casimir energy becomes positive within the range of value of
$\mu$, the ratio of plates distance to UXD size. From expression
(10), we find the relations between the special ratios $\mu_{1}$,
$\mu_{2}$ and temperature respectively shown in Figure 3.
$\mu_{1}$ and $\mu_{2}$ are increasing and decreasing functions of
scaled temperature respectively. If $\mu<\mu_{1}$ or
$\mu>\mu_{2}$, the Casimir energy will be negative. The nature of
the force between plates is attractive for $\mu<\mu_{1}$ and
repulsive for $\mu>\mu_{2}$. If $\mu_{1}<\mu<\mu_{2}$, the energy
will keep positive.

We can apply our results above to the approaches of string theory
with different size of UXDs. The environment temperature is chosen
as $T=300K$. The size of extra dimension is set $L~10nm$, the
suggestion of [24, 25], then $\xi\approx1.3\times10^{-3}<\xi_{0}$,
and $\mu_{f}\approx5.73$. The sign of Casimir energy is negative.
If the plates distance $R>50nm$, they repulse each other or the
opposite if $R<50nm$. We also study the model with
$L\sim10^{-2}cm$ similar to [5, 6]. Certainly
$\xi\approx13>\xi_{0}$, and if $\frac{R}{L}>3.53\times10^{9}$, the
nature of force between plates is repulsive while the Casimir
energy keeps negative. The result $R>3.53\times10^{5}m$ is
interesting.

\vspace{0.8cm}
\noindent 4.\hspace{0.4cm} Conclusion

In this letter we have explored the Casimir effect for the system
consisting of two parallel plates at finite temperature in the
Universe with extra dimensions. We derive the expression for the
Casimir energy with thermal corrections and one UXD. Having
studied the Casimir energy and Casimir force, we show that the
Casimir energy will remain negative if the temperature is less
than a special value. We also show that owing to the sufficiently
high temperature, the sign of Casimir energy will be positive when
the ratio of plates distance to UXD size chosen within the range.
The nature of Casimri force depends on the system structure and
environment. The distance between the parallel plates is larger or
less enough than the radius of UXD, their ratio more than or less
than a special value, then the Casimir energy keeps negative and
the plates repulse or attract each other. The two special ratios
of plates distance to the radius of extra space are increasing and
decreasing functions of the temperature respectively, which shows
that the Casimir effect is manifest if the temperature is high but
within our reach. According to our discussions on size of UXD from
models of string theory, the results that the plates repulse each
other under $R>3.53\times10^{5}m$ for $L\sim10^{-2}cm$ at
temperature $300K$ seem to be interesting.

\vspace{1cm}
\noindent Acknowledge

This work is supported by the Basic Theory Research Fund of East
China University of Science and Technology, grant No. YK0127312.

\newpage
\begin{figure}
\setlength{\belowcaptionskip}{10pt} \centering

  \caption{The solid, dadot, dashed curves of the Casimir energy density
  as functions of ratio of plates distance to UXD size for $\xi=0.094,0.0945,0.096$ respectively.}
\end{figure}

\newpage
\begin{figure}
\setlength{\belowcaptionskip}{10pt} \centering

  \caption{The curve of the special ratio of plates distance to UXD
  size $\mu_{f}$
  as a function of scaled temperature when $\xi<\xi_{0}$.}
\end{figure}

\newpage
\begin{figure}
\setlength{\belowcaptionskip}{10pt} \centering

\caption{The curves of the special ratios of plates distance to
UXD size for $\mu_{1}$ and $\mu_{2}$ as functions of scaled
temperature when $\xi>\xi_{0}$.}
\end{figure}

\end{document}